\documentclass[twocolumn,aps,prl]{revtex4}
\usepackage{amsmath}
\usepackage{amsfonts}
\usepackage{amssymb}
\usepackage{amsthm}
\usepackage{graphicx}

\begin{document}
\title{High field domain wall propagation velocity in 
magnetic nanowires}
%\title{Theory of field-induced domain wall propagation in 
%magnetic nanowires}
\author{X. R. Wang}
\affiliation{Physics Department, The Hong Kong University of
Science and Technology, Clear Water Bay, Hong Kong SAR, China}
\author{P. Yan}
\affiliation{Physics Department, The Hong Kong University of
Science and Technology, Clear Water Bay, Hong Kong SAR, China}
\author{J. Lu}
\affiliation{Physics Department, The Hong Kong University of
Science and Technology, Clear Water Bay, Hong Kong SAR, China}
%\date{\today}
\begin{abstract}
A thory of field-induced domain wall (DW) propagation is 
developed. The theory not only explains why a DW in a 
defect-free nanowire must propagate at a finite velocity, 
but also provides a proper definition of DW propagation 
velocity. This definition, valid for an arbitrary DW structure, 
allows one to compute the instantaneous DW velocity in a 
meaningful way even when the DW is not moving as a rigid body. 
A new velocity-field formula beyond the Walker breakdown 
field, which is in excellent agreement with both experiments 
and numerical simulations, is derived.
\end{abstract}
%\keywords{Domain-wall motion, magnetic nanowires}
% 75.60.Jk Magnetization reversal mechanisms\\
%75.75.+a Magnetic properties of nanostructures \\
%75.60.Ch Domain walls and domain structure  
%85.75.-d Magnetoelectronics; spintronics: 
%devices exploiting spin polarized transport or integrated magnetic fields  
%85.70.Ay Magnetic device characterization, design and modeling \\
\maketitle
%----------------------------------------------------------------%
It is a textbook knowledge\cite{book} that a magnetic field 
can drive a magnetic domain wall (DW) to move. However, 
our understanding of the field-induced DW motion is far 
from complete although it has been intensively studied for 
more than fifty years and many interesting phenomena of 
magnetization dynamics have been found. Recent development 
in nanomagnetism\cite{Parkin0} demands a deep understanding 
of DW motion in nanowires, especially how a field affects 
DW propagation velocity. DW dynamics is governed by the 
Landau-Lifshtiz-Gilbert (LLG) equation that can only be solved 
analytically for some special problems\cite{Walker,xrw}. 
A number of theories have been widely accepted and written 
in books\cite{book}, such as kinetic potential approach that 
assumes zero-damping, Thiele dynamic force equilibrium 
formulation that is correct for rigid DW propagation, Schryer 
and Walker analytical solution that is valid only for 1D and 
exact only for field smaller than a so-called Walker breakdown 
field $H_W$\cite{Walker}, Slonczewski formulation that 
simplifies a DW by its center and the cant angle of DW plane. 
None of these orthodox theories works beyond $H_W$ although 
they have greatly enriched our current understanding of DW 
dynamics. For example, kinetic potential approach cannot be 
a correct description of DW propagation because it violates 
the principle of ``no damping, no propagation" that will be 
explained in this paper. Thiele approach is a good way to 
describe a rigid DW propagation for small field $H<H_W$, but 
its assumptions are not valid for $H>H_W$. Schryer and 
Walker's approach is for 1D and $H<H_W$, and its predictions 
for $H>H_W$ are incorrect. For instance, its prediction that 
the $v-H$ line for $H>>H_W$ passing through the origin differs 
from both experiments and micromagnetic simulations\cite
{Ono,Cowburn,Erskine,Parkin1}. Its generalization predicts 
a saturated velocity\cite{Enz} (bounded by the velocity 
at $H_W$) that does not agree either with experiments 
or with simulations\cite{Ono,Cowburn,Erskine,Parkin1}. 
Slonczewski formulation is a great simplification of LLG 
equation that not only replaces partial differential equations 
by ordinary differential ones, but also is based on Thiele  
rigid DW approximation although the Slonczewski equations 
have also been applied to the case of $H>H_W$ where it is 
known that DW deformation cannot be neglected. The problems 
with both Thiele and Slonczewski formulations can also 
be seen from their $v-H$ formula\cite{Parkin1,Oleg} that do 
not capture the trend for $H>H_W$. Even more surprising, 
none of the existing theories provides a proper definition 
of DW propagation velocity when a DW does not propagate 
like a rigid body. 

In this paper, we develop a general formulation of DW 
propagation for both $H<H_W$ and $H>H_W$ that does not have 
all the problems with the existing theories. The theory 
reveals the origin of DW propagation. Firstly, we show 
that no static tail-to-tail (TT) or head-to-head (HH) DW 
is allowed in a homogeneous nanowire in the presence of an 
external magnetic field. A moving DW must dissipate energy 
because of various damping mechanisms. The energy loss 
should be supplied by the Zeeman energy released from the 
DW propagation. This energy consideration can clearly explain 
DW velocity oscillation for $H>H_W$. Secondly, the energy 
conservation provides a proper definition for DW propagation 
velocity. This definition leads to a general relationship 
between DW propagation velocity and the DW structure. 
Finally, a new velocity-field formula beyond the Walker 
breakdown field is derived and is compared with both 
experiments and numerical simulations.
\begin{figure}[htbp]
 \begin{center}
\includegraphics[width=7.cm, height=4.cm]{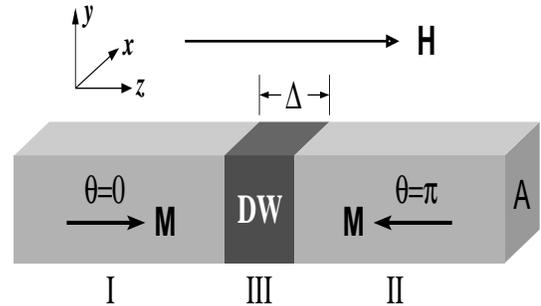}
 \end{center}
\caption{\label{fig1} Schematic diagram of a HH DW of width
$\Delta$ in a magnetic nanowire of cross-section $A$. 
The wire consists of three phases: two domains and one DW. 
The magnetization in domains I and II is along +z-direction 
($\theta=0$) and -z-direction ($\theta=\pi$), respectively. 
III is the DW region whose magnetization structure could 
be very complicate. $\vec H$ is an external field along 
+z-direction.} 
\end{figure}

In a magnetic material, magnetic domains are formed in order 
to minimize the stray field energy. A DW that separates two 
domains is defined by the balance between the exchange energy 
and the magnetic anisotropy energy. To describe a HH DW in a 
magnetic nanowire, let us consider a wire with its easy-axis 
along the wire axis which is chosen as the z-axis as 
illustrated in Fig. 1. Since the magnitude of the magnetization 
$\vec{M}$ does not change according to the LLG equation\cite{xrw},   
the magnetic state of the wire can be conveniently described by 
the polar angle $\theta(\vec x,t)$ (angle between $\vec M$ 
and the z-axis) and the azimuthal angle $\phi(\vec x, t)$. 
The magnetization energy of the wire energy can be written in 
general as 
\begin{equation}
\begin{split}
& E=\int F(\theta,\phi,\vec\nabla\theta,\vec\nabla\phi)d^3\vec x,\\ 
& F=f(\theta,\phi)+\frac{J}{2}[(\vec\nabla\theta)^2+\sin^2\theta
(\vec\nabla\phi)^2]-MH\cos\theta, 
\end{split}\label{energy}
\end{equation}
where $f$ is the energy density due to {\it all kinds} of 
magnetic anisotropies which has two equal minima at $\theta=0$ 
and $\pi$ ($f(\theta=0,\phi)=f(\theta=\pi,\phi)$). $J$ describes 
the exchange energy. $M$ is the magnitude of magnetization, 
and $H$ is the external magnetic field along z-axis. 
In the absence of $H$, a static HH DW that separates $\theta=0$ 
and $\theta=\pi$ domains can exist\cite{Walker} in the wire. 

{\it Non-existence of a static HH (TT) DW in a magnetic 
field}-In order to show that no intrinsic static HH DW is 
allowed in the presence of an external field ($H\neq 0$), 
one only needs to show that following equations have no solution 
with $\theta=0$ at far left and $\theta=\pi$ at far right, 
\begin{equation}
\begin{split}
&\frac{\delta E}{\delta\theta}=J\nabla^2\theta-\frac{\partial f}
{\partial\theta}-HM\sin\theta-J\sin\theta\cos\theta (\vec\nabla\phi)^2=0,\\
&\frac{\delta E} {\delta \phi}=J\vec\nabla\cdot(\sin^2
\theta \vec\nabla\phi)-\frac{\partial f}{\partial \phi}=0. 
\end{split}\label{DW}
\end{equation}
Multiply the first equation by $\nabla\theta$ and the second 
equation by $\nabla\phi$, then add up the two equations. 
One can show a tensor $\hbox{\bf T}$ satisfying 
$\nabla\cdot\hbox{\bf T}=0$ with 
\begin{align}
\hbox{\bf T}=& [-f+HM\cos\theta+\frac{J}{2}(|\nabla
\theta|^2+\sin^2\theta|\nabla\phi|^2)]\hbox{\bf 1}-\nonumber\\
& J(\nabla\theta\otimes \nabla\theta+
\sin^2\theta\nabla\phi\otimes \nabla\phi),\nonumber
\end{align}
where $\hbox{\bf 1}$ is $3\times 3$ unit matrix. 
$\nabla\theta\otimes\nabla\theta$ and $\nabla\phi\otimes\nabla
\phi$ are the usual dyadic products. The diagonal terms of 
$\hbox{\bf T}$ are just magentic Lagrangian density. 
If a HH DW exists with $\theta=0$ in the far left and $\theta=\pi$ 
in the far right, it requires $-f(0,\phi)+HM=-f(\pi,\phi)-HM$ that 
holds only for $H=0$ since $f(0,\phi)=f(\pi,\phi)$. 
In other words, a static DW can only exist between two 
equal-energy-density domains. A HH DW in a nanowire under 
an external field must vary with time because two domains 
separated by the DW have different magnetic energy density. 
It should be clear that the above argument is only true for 
a HH DW in a homogeneous wire, but not valid with defect 
pinning that changes Eq. \eqref{DW}. Static DWs do exist 
in the presence of a weak field in reality because of pinning. 

What is the consequence of the non-existence of a static DW? 
A DW has to move when an external magnetic field is applied 
along the nanowire as shown in Fig. 1. 
It is well known\cite{Thiv} that a moving magnetization must 
dissipate its energy to its environments with a rate, 
%\begin{equation}
$\frac{dE}{dt}=\frac{\alpha M}{\gamma}\int_{-\infty}
^{+\infty }\left(d\vec m/dt\right)^2d^3\vec x,$ 
%\end{equation}
where $\vec m$ is the unit vector of $\vec M$, $\alpha$ and 
$\gamma$ are the Gilbert damping constant and gyromagnetic 
ratio, respectively. Following the similar method in Ref.  
13 for a Stoner particle, one can also show that the energy 
dissipation rate of a DW is related to the DW structure as 
\begin{equation}\label{diss}
\frac{dE}{dt}=-\frac{\alpha\gamma}{(1+\alpha^2)M}\int_{-\infty}
^{+\infty }\left(\vec M\times\vec H_{eff}\right)^2d^3\vec x, 
\end{equation}
where $\vec H_{eff}=-\frac{\delta F}{\delta\vec M}$ is the 
effective field. In regions I and II or inside a static 
DW (Fig. 1), $\vec M$ is parallel to $\vec H_{eff}$,  
and no energy dissipation is possible there. 
The energy dissipation can only occur in the DW region 
when $\vec M$ is not parallel to $\vec H_{eff}$. 
%\begin{equation}\label{diss} 
%\frac{dE}{dt}=-\frac{\alpha\gamma}{1+\alpha^2}\int_{III}
%\left(\vec M\times \vec H_{eff}\right)^{2}d^3\vec x. 
%\end{equation}

{\it DW propagation and energy dissipation-}For a magnetic 
nanowire in a static magnetic field, the dissipated energy must 
come from the magnetic energy released from the DW propagation. 
The total energy of the wire equals the sum of the energies of 
regions I, II, and III (Fig. 1), $E=E_I+E_{II}+E_{III}$. 
$E_I$ increases while $E_{II}$ decreases when the DW 
propagates to the right along the wire. The net energy change 
of region I plus II due to the DW propagation is 
%\begin{equation}\label{diss1}
$\frac{d(E_{I}+E_{II})}{dt}=-2HMvA,$
%\end{equation}
where $v$ is the DW propagating speed, and $A$ is the area of 
wire cross section. This is the released Zeeman energy stored in 
the wire. The energy of region III should not change much because 
DW width $\Delta$ is finite, typically order of $10\sim 100nm$. 
A DW cannot absorb or release too much energy, 
and can at most adjust temporarily energy dissipation rate. 
In other words, $\frac{dE_{III}}{dt}$ is either zero or fluctuates 
between positive and negative values with zero time-average. 
Since energy release from the magnetic wire should be equal to 
the energy dissipated (to the environment), one has 
%\begin{equation}
%-2HMvA +\frac{dE_{III}}{dt}=-\frac{\alpha\gamma}{(1+\alpha^2)M}
%\int_{III}\left(\vec M\times \vec H_{eff}\right)^{2}d^3\vec x. 
%\end{equation}
%or 
\begin{equation}\label{main}
v= \frac{\alpha\gamma }{2(1+\alpha^2)HA}\int_{III}\left(\vec m\times 
\vec H_{eff}\right)^{2}d^3\vec x +\frac{1}{2HMA}\frac{dE_{III}}{dt}.
\end{equation}
Eq. \eqref{main} can serve as a proper definition of DW propagation 
velocity that is completely defined by the {\it instantaneous} 
DW structure. 

{\it Velocity oscillation-}A DW can have two possible types 
of motion under an external magnetic field. One is that a DW 
behaves like a {\it rigid body} propagating along the wire. 
This case occurs often at low field, and it is the basic 
assumption in Slonczewski model\cite{Slon} and Walker's 
solution for $H<H_W$. Obviously, both energy-dissipation 
and DW energy is time-independent, $\frac{dE_{III}}{dt}=0$. 
Thus, the DW velocity should be time-independent. 
The other case is that a DW structure varies with time which 
occurs at large field $H>H_W$. In this case, DW structure 
deforms and DW precesses around the wire axis, experiencing 
different transverse magnetic anisotropy energy. As a result, 
the DW width breathes periodically since it is defined by 
the balance between the magnetic anisotropy energy and the 
exchange energy. Thus, one should expect both
$\frac{dE_{III}}{dt}$ 
and energy dissipation rate oscillate with time. 
According to Eq. \eqref{main}, DW velocity will oscillate. 
DW velocity may oscillate periodically or irregularly, 
depending on whether the ratio of precession period and 
breathe period is rational or irrational. Indeed, this 
oscillation was observed in a recent experiment\cite{Erskine}. 
How can one understand the wire-width dependence of DW velocity? 
According to Eq. \eqref{main}, the velocity is a functional 
of DW structure which is very sensitive to the wire width. 
For a very narrow wire, only transverse DW is possible while a 
vortex DW is preferred for a wide wire (large than DW width). 
Different vortexes yield different values of $|\vec m\times 
\vec H_{eff}|$, which in turn results in different DW 
propagation speed. 
\begin{figure}[htbp]
 \begin{center}
\includegraphics[width=9.5cm, height=11.5cm]{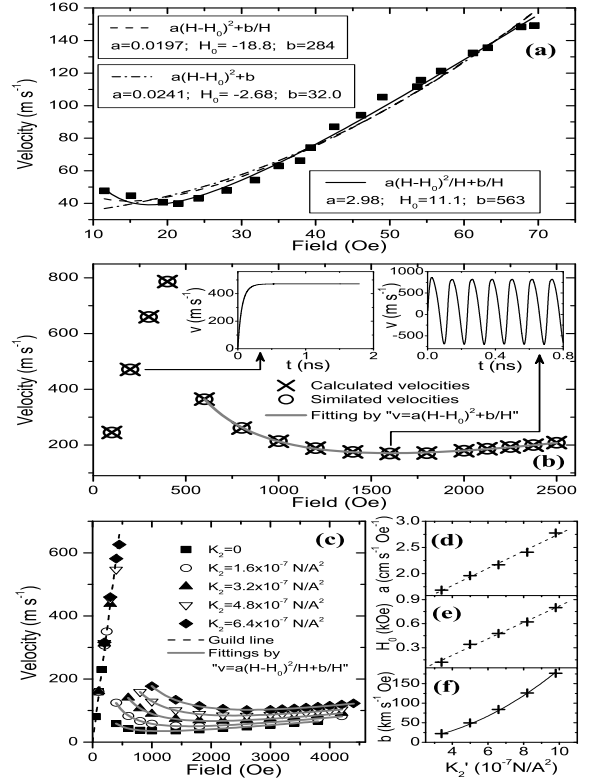}
 \end{center}
\caption{\label{fig2} a) Symbols are from Ref. 8a. Solid 
line is the best fit to Eq. (8). Dash and dash-dot lines 
are fits to some modified expressions of Eq. (8). b) The 
time-averaged DW propagation speed versus the applied 
magnetic field for a biaxial magnetic nanowire. The cross 
are calculated velocities from Eq. \eqref{ave}, and the 
open circles are simulated average velocities. 
The solid curve is the fit to Eq. (8). Insets: the 
instantaneous DW speed calculated from Eq. \eqref{main} 
for $H=50Oe <H_W$ (left) and $H=1000Oe>H_W$ (right). 
c) Symbols are numerical simulations of another set of 
parameters. The dashed line is the linear fit for low 
field and solid lines are the fits to Eq. (8). d)-f) 
The dependence of $a$, $H_0$, and $b$ on $K_2'$ defined 
in the text. 
%\equiv K_2 +3.4\times 10^{-7} N/A^2$, where 
%$3.4\times 10^{-7} N/A^2$ is due to the demagnetization 
%effect obtained by the theory in Ref. 15. 
Dash lines are linear fits, and solid line is quadratic fit.} 
\end{figure}

Time averaged velocity is 
\begin{equation}\label{ave}
\bar v= \frac{\alpha\gamma}{2(1+\alpha^2)HA}\int_{III}
\overline{\left({\vec m\times \vec H_{eff}}\right)^{2}}d^3\vec x, 
\end{equation}
where bar denotes time average. It says that the averaged velocity 
is proportional to the energy dissipation rate. 
The right-hand side of Eq. \eqref{ave} is positive and non-zero 
since a time dependent DW requires $\vec m\times\vec H_{eff}\neq 0$,
implying a zero intrinsic critical field for DW propagation.
It is straightforward to show 
\begin{equation}
\begin{split}
& \left(\vec m\times \vec H_{eff}\right)^{2}=
\left\lbrace \frac{1}{M\sin\theta}\left[ J\frac{\partial}{\partial z}
\left(\sin^2\theta \frac{\partial \phi}{\partial z}\right) 
-\frac{\partial f}{\partial\phi}\right]\right\rbrace^2 + \\
&\left\lbrace H\sin\theta-\frac{1}{M}\left[ J\frac{\partial^
2\theta}{\partial z^2}-\frac{\partial f}{\partial\theta}-J
\sin\theta\cos\theta (\frac{\partial\phi}{\partial 
z})^2\right] \right\rbrace^2. 
\end{split}\label{product}
\end{equation}
It is interesting to notice that two terms on the right hand side 
of Eq. \eqref{product} are just $\left[\frac{1}{M}\frac{\delta E}
{\delta\phi}\right ]^2$ and $\left[\frac{1}{M\sin\theta}\frac
{\delta E}{\delta\theta}\right ]^2$. In the case that there is no 
distortion on the DW plane, ie. $\partial\phi/\partial z=0$ and
$\partial^2\phi/\partial z^2=0$, then 
\begin{equation}\label{product1}
\left({\vec m\times \vec H_{eff}}\right)^{2}=(H-H_0^\prime)^2\sin^2\theta+
\left(\frac{1}{M\sin\theta}\frac{\partial f}{\partial\phi}\right)^2
\end{equation} 
where $H_0^\prime\equiv \frac{1}{M\sin\theta}\left(J\frac{\partial^2
\theta}{\partial z^2}-\frac{\partial f}{\partial\theta}\right).$ 
To reproduce the result of Schryer and Walkeri\cite{Walker} 
for $H<H_W$, one sets 
$\frac{\partial \phi}{\partial z}=0$, $\frac{\partial \phi}
{\partial t}=0$, and $J\frac{\partial^2\theta} {\partial z^2}
=\frac{\partial f}{\partial\theta}$ (similar to the original 
analysis of Schryer and Walker\cite{Walker}) so that $H_0^\prime=0$.
Under the assumption, LLG Eq. gives $\alpha\frac{\partial\theta}
%in terms of $\theta$ and $\phi$ leads to 
{\partial t}=\gamma H \sin\theta$ and $\frac{\partial \theta}
{\partial t}= -\frac{\gamma}{M\sin\theta}\frac{\partial f} 
{\partial \phi}$. Thus $\frac{1}{M\sin\theta}\frac{\partial f} 
{\partial \phi}=-\frac{H}{\alpha}\sin\theta$, and the right-hand 
side of Eq \eqref{product1} is just $\frac{1+\alpha^2}{\alpha^2} 
H^2\sin^2\theta$. Using the definition of DW width $\Delta$, 
$\int \sin^2\theta d^3\vec x = 2\Delta A$, then Eq. \eqref{ave} 
reproduces the famous  Walker's mobility 
coefficient $\mu=\frac{\gamma\Delta}{\alpha}$ for $H< H_W$. 
For $H>H_W$, $\phi$ varies periodically with time between $0$ 
and $2\pi$, Substitute expression in Eq. \eqref{product} into 
Eq. \eqref{ave} and take the time average, field dependent 
averaged velocity takes a form of $\bar v=aH-a_{0}+a_{-1}/H$ 
that can be rewritten as 
\begin{equation}\label{v-h}
\bar v= a(H-H_0)^2/H+b/H,
\end{equation}
where $a$ is proportional to the averaged DW width, $H_0$ and 
$b$ (and $a_0$, $a_{-1}$) depend on DW structure and magnetic 
anisotropy. 

To demonstrate the goodness of Eq. \eqref{v-h}, we fit the 
experiment data (symbols in Fig. 2a) from Ref. 8a by the 
expression (solid line in Fig. 2a) with $a=2.98m/(s\cdot Oe)$, 
$b=563 Oe\cdot m/s$, and $H_0=11Oe$. The experimental mobility 
at large field is measured to be $2.5 m/(s\cdot Oe)$ that 
compares well with $a=2.98m/(s\cdot Oe)$. According to our 
theory, $b$ should be proportional to $\alpha\gamma\bar\Delta 
K_2^2/(1+\alpha^2)$ ($K_2$, defined later, measures the 
transverse magnetic anisotropy (TMA)), where $\bar\Delta$ is 
the time-averaged DW width. Using material parameters\cite
{Erskine} $M=860\times 10^3 A/m$; $J=13\times 10^{-12}J/m$; 
$K_1M^2=500 J/m^3$ (defined later); $\alpha=0.01$, and 
measured low-field mobility $\mu=25m/(s\cdot Oe)$; and Walker 
breakdown field $H_W=4Oe$, the DW width for $H<H_W$ is about 
$\Delta=14nm$, and TMA constant to be $K_2M=34\times 10^3A/m$ 
from $\Delta=\mu\alpha/\gamma$ and $H_W=\alpha K_2M$, the 1D 
result from Walker's original paper. It is known that DW width 
should vary as the DW precess around wire axis for $H>H_W$. 
Although exact value of the averaged width $\bar\Delta$ is 
not known from the experiment, the theory requires its value 
to be about $16nm$ in order to obtain the  fitting value of 
$b=563 Oe\cdot m/s$. $16nm$ is a fair value for DW width. 
The good agreement between Eq. \eqref{v-h} and experimental 
results is not accidental. In fact, if one changes the fitting 
formula slightly to $a(H-H_0)^2+b/H$ or $a(H-H_0)^2+b$, the 
best fits (dashed line and dashed-dot line in Fig. 2a) not only 
show poor agreement with experiment, but also give unreasonable 
fitting parameters. This proves that a good fitting is not due 
to three fitting parameters introduced.

To further test the validity of Eq. \eqref{v-h} and usefulness 
of both Eqs. \eqref{main} and \eqref{ave} in evaluating the DW 
propagation speed from a DW structure, we carry out 
micromagnetic simulations on a strap wire of $4nm\times 20nm
\times 3\mu m$ whose magnetic energy density is $F=-\vec M\cdot
\vec H+\frac{J}{M^2}(\nabla\vec M)^2 -K_1 M_z^2+K_2M_x^2$. 
We use OOMMF package\cite{oommf} to find the DW structures and 
then use Eq. \eqref{ave} to obtain the average velocity or Eq. 
\eqref{main} for instantaneous velocity. 
Fig. 2b is the simulation results for system parameters of 
$K_1=10^{-7}N/A^2$, $K_2=0.8\times 10^{-7}N/A^2$, 
$J=4\times 10^{-11}J/m$, $M=10^6A/m$, and $\alpha=0.1$. 
The calculated velocities are denoted by cross and numerical 
simulations are the open circles with their error bars smaller 
than the symbol sizes. The good overlap between the cross and 
open circles confirm the correctness of Eq. \eqref{ave}. 
The $\bar v -H$ curve for $H>H_W$ can be fit well by  Eq. 
\eqref{v-h}. The insets are instantaneous DW propagation 
velocities for both $H<H_W$ and $H>H_W$, by Eq. \eqref{main} 
from the instantaneous DW structures obtained from OOMMF. 
It should be emphasized that both $\phi$ and $\theta$ are 
complicated functions of the time and the space.
The left inset is the instantaneous DW speed 
at $H=50Oe<H_W$, reaching its steady value in about $1ns$. 
The right inset is the instantaneous DW speed at $H=1000Oe
>H_W$, showing clearly an oscillation. They confirm that the 
theory is capable of capturing all the features of DW 
propagation. 

Figure 2c is another set of OOMMF simulations on a wire of 
$4nm\times 16nm \times 4 \mu m$ with system parameters of 
$M=500 kA/m$ for the saturation magnetization, $J=20\times 
10^{-12}J/m$, the axial crystalline anisotropy constant 
$K_1=8\times 10^{-7}N/A^2$, and  $\alpha=0.1$. The symbols 
are simulation results, and solid lines are the fits to Eq. 
\eqref{v-h}. The perfect agreement demonstrate the correctness 
of Eq. \eqref{ave}. Fig. 2d-2f show how $a$, $b$ and $H_0$ 
depend on the normalized TMA constant $K'_2=K_2+3.4\times 10
^{-7}N/A^2$. This normalized parameter comes from the 
demagnetization effect that generates an extra magnetic 
anisotropic energy $(D_1-D_2)M_x^2+(D_3-D_2)M_z^2+D_2$, 
where $D_i$ is the demagnetization factor along i-th axis. 
Using the theory in Ref. 15, one can find $D_1-D_2$ to be 
$3.4\times 10^{-7}N/A^2$ for our geometry. $a$ and $H_0$ are 
linear in $K_2'$ while $b$ is quadratic in $K_2'$, in 
agreement with Eq. (7). It should be pointed out that velocity 
expressions in both Refs. 9 and 10 cannot fit either the 
experimental curve (Fig. 2a) or simulations (Figs. 2b and 2c). 
The correctness of result Eq. \eqref{main} depends only on 
the LLG equation and the general energy expression of Eq. 
\eqref{energy}. It does not depend on the details of a DW 
structure whether they are transverse or vortex like. 
In this sense, our result is very general and robust, 
and it is applicable to an arbitrary magnetic wire. 

In conclusion, a proper definition of DW propagation 
velocity is obtained, and a velocity-field formula for 
high field (above the Walker breakdown field) is proposed. 
This new formula agrees well with both experiments and 
numerical simulations. Furthermore, a global picture of DW 
propagation in a nanowire driven by a magnetic field is 
revealed: A static DW cannot exist in a homogeneous magnetic 
nanowire when an external magnetic field is applied. 
Thus, a DW must vary with time under a static magnetic field. 
A moving DW must dissipate energy due to the Gilbert damping. 
As a result, the wire has to release its Zeeman energy through 
the DW propagation along the field direction. The DW propagation 
speed is proportional to the energy dissipation rate that is 
determined by the DW structure. An oscillatory DW motion, 
either the precession around the wire axis or the breath 
of DW width, should lead to the speed oscillation. 
The observed negative differential mobility is due to the 
transition of a DW from a high energy dissipation state to 
a low energy dissipation one.

This work is supported by Hong Kong UGC/CERG grants 
(\# 603007, 603508, and SBI07/08.SC09).

\end{document}